\documentclass[aps,%
prl,%
twocolumn%
]{revtex4}

\usepackage[dvips]{graphicx}

\def\figtag{Fig. }
\def\figstag{Figs. }

\def\refstag{Refs. }

\newcommand{\thermav}[1]{\langle #1 \rangle}

\renewcommand{\vec}[1]{\mbox{{\boldmath$#1$}}}

\def\gsim{\buildrel {\textstyle >}\over {_\sim}}

\begin{document}
\title{Replica symmetry breaking transition of the weakly
anisotropic Heisenberg spin glass
in magnetic fields}
\author{Daisuke Imagawa and Hikaru Kawamura}
\affiliation{Department of Earth and Space Science, Faculty of Science,
Osaka University, Toyonaka 560-0043,
Japan}
\date{\today}
\begin{abstract}
The spin and the chirality orderings of the three-dimensional Heisenberg
spin glass with the weak random anisotropy are studied
under applied magnetic fields by  equilibrium Monte Carlo simulations. 
A replica symmetry breaking
transition occurs in
the chiral sector accompanied by the simultaneous spin-glass order.
The ordering behavior differs significantly from that of the 
Ising SG, despite the similarity in the global symmetry.
Our observation is consistent 
with the spin-chirality decoupling-recoupling scenario of a spin-glass
transition. 
\end{abstract}
\maketitle

Whether or not spin-glass (SG) magnet exhibits a thermodynamic phase transition
in applied magnetic fields has been a long-standing issue\cite{Review}.
This issue is closely
related to the fundamental question of whether 
the SG ordered state in zero field
accompanies an ergodicity breaking not directly related
to the global symmetry of the Hamiltonian, {\it i.e.\/}, the
replica symmetry breaking (RSB).
Most of numerical studies on SG have focused on the properties of
the Ising SG\cite{Review}. Since the Ising SG
possesses no global symmetry in magnetic fields, the
occurrence of a phase transition in fields
would necessarily mean the occurrence of RSB.
Unfortunately, numerical simulations on
the Ising SG 
have been unable to give a 
definitive answer concerning the existence of
a SG transition in magnetic fields\cite{Picco,Cruz}.

Experimentally,
some evidence against an
in-field transition has been reported
for the strongly anisotropic
Ising-like SG Fe$_{0.5}$Mn$_{0.5}$TiO$_3$\cite{Nordblad}.
In fact, many of
real SG materials are more or less Heisenberg-like rather than Ising-like,
in the sense that the magnetic anisotropy is considerably weaker than
the isotropic exchange interaction\cite{Review}. 
Recent experiments on such
weakly anisotropic Heisenberg-like SG
suggested the occurrence of an in-field SG transition\cite{Campbell}, 
in apparent contrast to Ref.\cite{Nordblad}.

Meanwhile,  via recent theoretical studies,
it has become increasingly clear 
that the Heisenberg SG
possesses an important physical ingredient which is totally
absent in the Ising SG, {\it i.e.\/}, 
{\it the chirality\/}\cite{Kawa92,Kawa98}.
In particular,
the chirality scenario 
of Ref.\cite{Kawa92,Kawa98} claims that the chirality is
a hidden order parameter of the
SG transition of real Heisenberg-like SG magnets. According to this
spin-chirality decoupling-recoupling scenario, in the fully isotropic
Heisenberg SG, the spin and the chirality, which are coupled at short
length/time scales, are eventually decoupled at long
length/time scales, and
the system exhibits a chiral-glass transition without  the
standard SG order. In the weakly anisotropic Heisenberg SG,
the Heisenberg spin
is ``recoupled'' to the chirality
at these long length/time scales via the random magnetic anisotropy.
The SG order of the weakly anisotropic Heisenberg SG is then dictated
by the ordering of the chirality. In zero field,
some numerical support for such a scenario
was reported\cite{HukuKawa,KawaLi,ImaKawa2,KawaYone,MatsuHuku},
although some other groups claimed that the chiral-glass
transition of the 
isotropic system already accompanied the SG order
\cite{Nakamura,Young}.

In connection to experiments,  
in-field ordering properties of the Heisenberg SG, both with and without
the random magnetic anisotropy, are of obvious interest.
The recent Monte Carlo (MC) simulation
of the fully isotropic Heisenberg SG
revealed that
the chiral-glass transition corresponding to the breaking of the chiral
$Z_2$ symmetry occurred even under fields, while the
observed transition line
had a striking resemblance to the so-called Gabay-Toulouse
line observed experimentally\cite{ImaKawa}.
Note, however, that the fully isotropic Heisenberg SG
possesses 
$Z_2\times SO(2)$ symmetry even  in magnetic fields, 
the chiral $Z_2$ corresponding to
a global spin reflection with respect to the plane containing the
magnetic-field axis and the $SO(2)$ to
the global spin rotation around the magnetic-field axis. 
In the more realistic case of the weakly anisotropic Heisenberg SG,
there no longer remains any global symmetry under fields. Hence,
from symmetry, the
situation is common with that of the Ising SG with the infinitely
strong anisotropy. Meanwhile,
the Heisenberg SG possesses the
nontrivial chiral degrees of freedom, which are
totally absent in the Ising SG.

Under such circumstance, it would be highly interesting to
examine the ordering properties of the
weakly anisotropic three-dimensional (3D) Heisenberg SG 
in magnetic fields as compared with 
those of the
Ising SG. Such an interest is further promoted by the 
apparently  contradicting experimental observations on the Ising-like and
the Heisenberg-like SGs\cite{Nordblad,Campbell}.
In the present Letter,  we study  by extensive MC simulations 
the spin-glass and the chiral-glass
orderings of the weakly anisotropic Heisenberg SG in magnetic fields.

The model considered is the anisotropic classical Heisenberg
model on a 3D simple cubic lattice defined by the
Hamiltonian,
\begin{equation}
{\cal H}=-\sum_{\thermav{ij}}(
J_{ij}\vec{S}_i\cdot\vec{S}_j
+ \sum_{\mu,\nu=x,y,z}D_{ij}^{\mu\nu}S_{i\mu}S_{j\nu})
- H\sum_{i=1}^{N} S_{iz},
\label{hamil}
\end{equation}
where $\vec{S}_i=(S_{ix},S_{iy},S_{iz})$ is a three-component unit vector,
and $H$ is the intensity of the magnetic field applied along
the $z$ direction.
The isotropic nearest-neighbor exchange coupling 
$J_{ij}$ is assumed to take either
the value $J$ or $-J$ with equal probability
while the  nearest-neighbor random exchange anisotropy
$D_{ij}^{\mu\nu}$'s ($\mu,\ \nu$=$x,y,z$ are spin-component indices) 
are assumed to be
uniformly distributed in the range $[-D:D]$,
where $D$ is the typical intensity of the anisotropy.
We impose the relation
$D_{ij}^{\mu\nu}=D_{ji}^{\mu\nu}=D_{ij}^{\nu\mu}$.

Simulations are performed for the anisotropy
$D/J=0.05$ and for
a variety of field intensities in the range
$H/J=[0.02$:$5.0$], whereas  the results shown below are for
the field intensity $H/J=0.05$ at which
most extensive calculations are made.
The lattice contains $L^3$ sites
with $L=4$, 6, 8, 10, 12, 16  and 20 with periodic boundary conditions.
Sample average is taken over 40-600 independent bond realizations.
To facilitate efficient thermalization, we combine the standard
heat-bath method  with the temperature-exchange technique
~\cite{TempEx}. Following the procedure of Ref.\cite{ImaKawa2},
care is taken to be sure that the system is fully equilibrated.

The local chirality, $\chi_{i\mu}$, is defined
at the $i$-th site and in the $\mu$-th
direction for three neighboring Heisenberg spins
by the scalar, $\chi_{i\mu}=
\vec{S}_{i+\hat{\vec{e}}_{\mu}}\cdot
(\vec{S}_i\times\vec{S}_{i-\hat{\vec{e}}_{\mu}})$, 
%
%
where $\hat{\vec{e}}_{\mu}\ (\mu=x,y,z)$ denotes a unit vector along the
$\mu$-th axis. 
By considering two independent systems (``replicas'') described by
the same Hamiltonian (\ref{hamil}),
we define the overlaps of the chirality and of the spin by,
\begin{eqnarray}
q_{\chi}
&=&
\frac{1}{3N}\sum_{i,\mu}
\chi_{i\mu}^{(1)}\chi_{i\mu}^{(2)}\ \ ,
\end{eqnarray}
\begin{eqnarray}
q_{\mu\nu}
&=&
\frac{1}{N}\sum_{i=1}^N S_{i\mu}^{(1)}S_{i\nu}^{(2)}, 
\end{eqnarray}
where $(1)$ and $(2)$ denote the replicas 1 and 2.
In our simulations, the two replicas 1 and 2 are prepared by
running two independent sequences of  systems
in parallel with different spin initial conditions and
different sequences of random numbers.

We first compute the equilibrium time-correlation functions of the chirality
and of the $z$-component of the Heisenberg spin
(henceforth simply  denoted as ``spin" ), 
%
%
%
\begin{eqnarray}
q^{(2)}_{\chi}(t)
&=&
\left[
\left\langle
\left(\frac{1}{3N}\sum_{i,\mu}^N\chi_{i\mu}(t_0)\chi_{i\mu}(t+t_0)\right)^2
\right\rangle
\right]
-[\thermav{q_{\chi}^2}]
\ \ ,
\label{Cxt}
\\
q^{(2)}_{z}(t)
&=&
\left[
\left\langle
\left(\frac{1}{N}\sum_{i=1}^NS_{iz}(t_0)S_{iz}(t+t_0)\right)^2
\right\rangle
\right]
-[\thermav{q_{zz}^2}]\ \ ,
\label{Czt}
\end{eqnarray}
where  $\thermav{\cdots}$ represents the thermal average and
[$\cdots$] the average over the bond disorder, while
the ``time'' $t$ is measured in units of MC sweeps.
Note that, in the above definitions,
the second terms $[\thermav{q_\chi^2}]$
and $[\thermav{q_z^2}]$ have been subtracted, 
which are nonzero even in the high-temperature phase in the
$L\rightarrow \infty$ limit due to the
absence of the global symmetry.
This subtraction
guarantees that both $q^{(2)}_\chi(t)$ and $q^{(2)}_z(t)$ decay to zero as
$t\rightarrow \infty$  in the
high-temperature phase. In the possible ordered phase,
by contrast,
both $q^{(2)}_\chi(t)$ and $q^{(2)}_z(t)$ decay to zero {\it if\/} the ordered
state does not
accompany the RSB, but tend to finite positive values {\it if\/} the
ordered state accompanies the RSB.
The latter property arises because  in the presence of RSB 
the $t\rightarrow \infty$ limits of the first terms are generally greater than
$[\thermav{q_\chi^2}]$ and $[\thermav{q_z^2}]$.
In computing the first terms,
we  perform simulations
according to the standard heat-bath updating without the
temperature-exchange procedure where
the starting spin configuration at $t=t_0$ is taken from
the equilibrium spin configurations
generated in our temperature-exchange MC runs, while the seconds terms
are evaluated in the temperature-exchange MC runs.

\begin{figure}[ht]
  \leavevmode
  \begin{center}
    \begin{tabular}{l}
      \includegraphics[width=\linewidth]{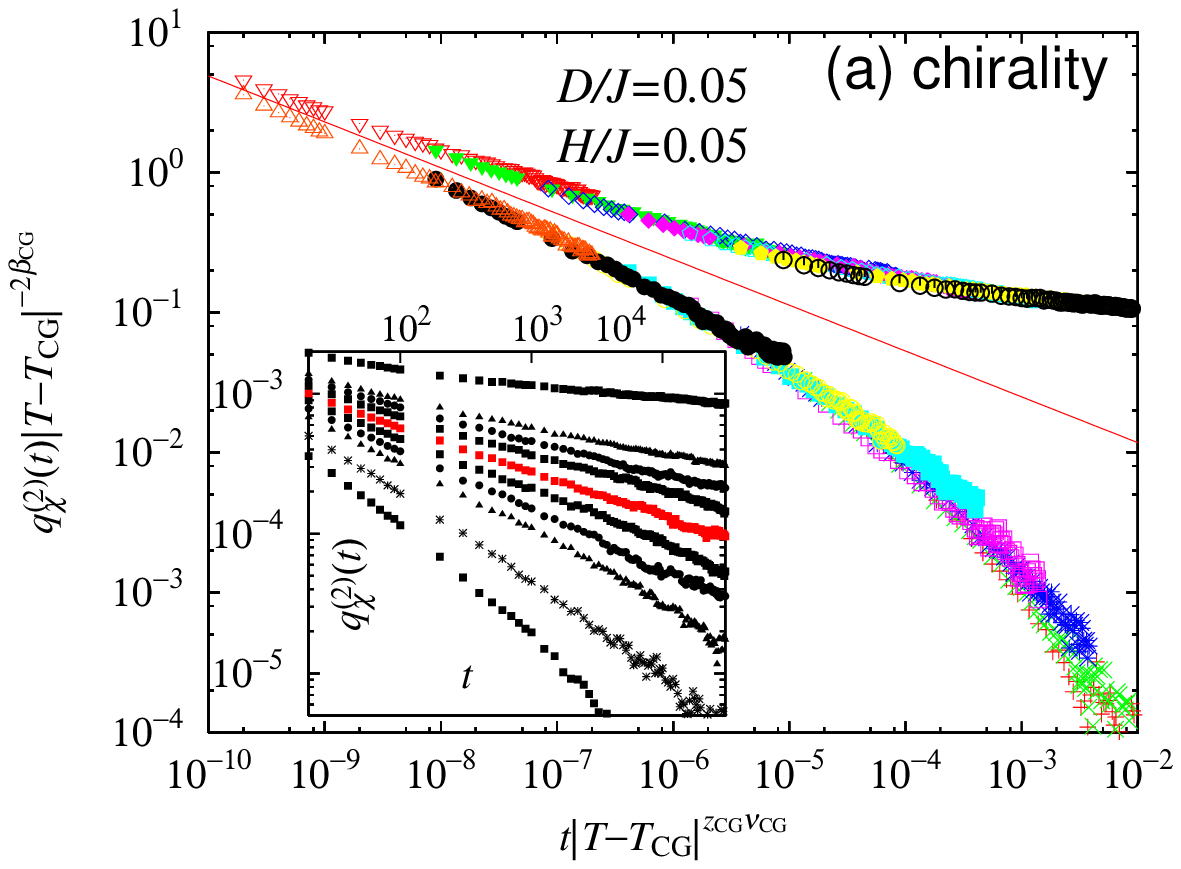}
    \end{tabular}
    \begin{tabular}{l}
      \includegraphics[width=\linewidth]{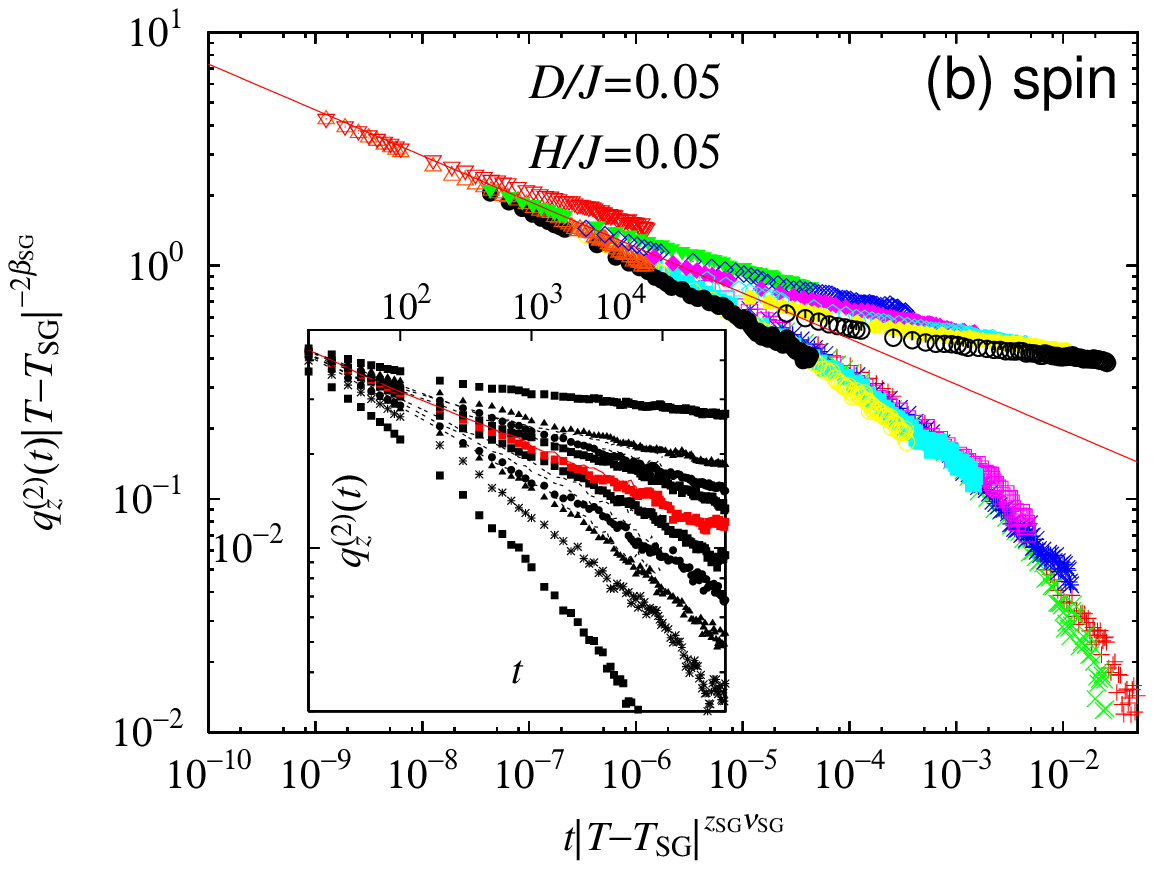}
    \end{tabular}
\caption{
Temporal decay of the equilibrium time-correlation 
functions of (a) the chirality defined by eq.(4), and
(b) the spin defined by eq.(5),
for $D/J=0.05$ and $H/J=0.05$.
The lattice size is $L=16$.
The data at $T=0.14$, those in the temperature range $T=
0.18\sim 0.24$ with an interval of 0.01, and those at $T=0.26$ and 0.28
are plotted 
(from above to below in the insets). The estimated transition
temperature is $T_g\simeq 0.21$. Main panels represent the dynamical
scaling plots, where the exponents are taken to be
(a) $\beta_{{\rm CG}}=0.9$ and $z_{{\rm CG}}\nu_{{\rm CG}}=5.5$,
while (b) $\beta_{{\rm SG}}=0.5$ and $z_{{\rm SG}}\nu_{{\rm SG}}=5.1$.
Slope of the straight dashed line is
equal to $2\beta/z\nu$. Insets represent the raw data, where
the data at $T=T_g$ are given in red.}
    \label{fig_Ct}
  \end{center}
\end{figure}

The chiral and the spin time-correlation functions  
of the size $L=16$
are shown on log-log plots in \figstag\ref{fig_Ct} (a) and (b), respectively. 
A comparison with the $L=20$ data indicates that
the data can be regarded as those of the bulk  in the time range shown,
since no appreciable size effect is discernible either in the first or the
second terms of eqs. (4) and (5).
As shown in the insets, both $q^{(2)}_\chi(t)$ and $q^{(2)}_z(t)$ 
exhibit either
down-bending or up-bending behavior depending on whether the temperature is
higher or lower than $T\simeq 0.21$ (in units of $J$),
while just at $T\simeq 0.21$
straight-line behavior corresponding a power-law decay is observed.
This clearly indicates that both the spin and the chirality exhibit a phase
transition at $T=T_g\simeq 0.21$ into the low-temperature phase where the
replica symmetry
is broken. A simultaneous ordering of the spin and the chirality seems
quite natural in the presence of the random anisotropy. The estimated
transition temperature $T_g=0.21\pm 0.02$ is lower than $T_g$ in  zero field,
$T_g\simeq 0.24$\cite{MatsuHuku, HukuKawa3},
indicating that the applied field suppresses the chiral-glass
(spin-glass) ordering for weak fields.
This suppression of $T_g$
due to  applied fields is
consistent with the experimental observation\cite{Campbell}.
We  also calculate
the transverse (perpendicular to the applied field) spin
time-correlation function, observing the qualitatively
similar behavior as $q^{(2)}_\chi$ and $q^{(2)}_z$ (the data
not shown here).

%
%

We then perform a dynamical scaling analysis both
for $q^{(2)}_\chi(t)$ and $q^{(2)}_z(t)$,
with setting $T_g=0.21$. As shown
in \figtag\ref{fig_Ct} (a), the chiral time-correlation function
scales very well  both above and below
$T_g$  with $\beta_{{\rm CG}}=0.9$ and $z_{{\rm CG}}\nu_{{\rm CG}}=5.5$,
where $\beta_{{\rm CG}}$,
$\nu_{{\rm CG}}$ and
$z_{{\rm CG}}$ refer to the order
parameter, the correlation-length and the dynamical chiral-glass exponents,
respectively.
The estimated chiral-glass exponents are
not far from the corresponding zero-field values\cite{MatsuHuku}.
By contrast, the spin time-correlation function
scales not quite well for any choice of the fitting parameters:
An apparent ``best fit'' is given in
\figtag\ref{fig_Ct} (b), whereas the quality of the fit is not improved for other
choices. Such a poor scaling suggests that, {\it for the
spin\/}, the asymptotic
scaling regime has not been reached in the investigated time range
and that
the fitted values of $\beta_{{\rm SG}}$ and $z_{{\rm SG}}\nu_{{\rm SG}}$
might not be true asymptotic values.
The reason
why the spin time-correlation exhibits a poorer scaling even in the
regime where
the chirality exhibits a nice scaling may naturally be understood
from the spin-chirality decoupling-recoupling scenario:
In this scenario, the chirality is always the order parameter
of the transition, while
the spin, decoupled from the chirality
in the absence of the anisotropy
at long length/time scales (estimated to be $r\gsim 20$
and $t\gsim 10^5$),
is recoupled to the chirality in the
presence of the anisotropy at these long scales and eventually shows
essentially the same ordering behavior as the chirality, $\chi\approx S$.
At shorter scales, by contrast, the spin and the chirality are
trivially coupled by its definition, roughly being $\chi\approx S^3$,
irrespective of the anisotropy.
If so, an asymptotic critical behavior {\it of the spin\/}
would be evident 
only at longer times $t\gsim 10^5$, which is beyond the time range
of \figtag\ref{fig_Ct}.

%
\begin{figure}[ht]
  \begin{center}
    \includegraphics[width=\linewidth]{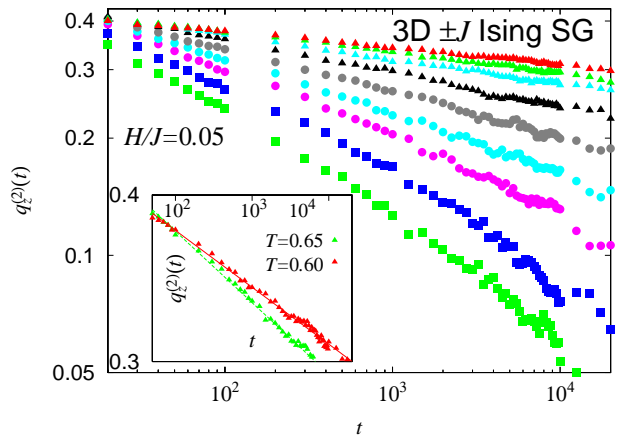}
\caption{Temporal decay of the equilibrium time-correlation
function of the spin, defined by eq.(5),
of the $\pm J$ 3D Ising SG in a magnetic field of $H/J=0.05$.
The system size is $L=16$ averaged over 125 samples.
The data in the temperature range
$T=0.60\sim 1.4$  with an interval of 0.1 are plotted in the
main panel from above 
to below. 
The data of the two lowest temperatures $T=0.60$ and 0.65
are shown in the inset,
together with the fitted straight lines. At any temperature
studied, no up-bending behavior is observed.
}
    \label{fig_Ct_3DISG}
  \end{center}
\end{figure}

For comparison, we also calculate the spin time-correlation
function (\ref{Czt}) for the 3D Ising SG with the $\pm J$ coupling for the
field $H/J=0.05$. The
result for $L=16$ is shown in \figtag\ref{fig_Ct_3DISG}.
Here, the behavior of $q^{(2)}_z(t)$
differs significantly from that of the weakly anisotropic
Heisenberg SG:
Although the temperature range studied is as low
as about 60\%
of the zero-field transition temperature $T_g(H=0)\simeq 1.1$, which
is expected to be deep in the ordered state 
according to a tentative estimate of \refstag\cite{Cruz},
no clear up-bending behavior as observed in  the weakly anisotropic
Heisenberg SG is observed. Instead,
$q^{(2)}_z(t)$ at lower temperatures persistently exhibits
an almost linear behavior. 
This is illustrated
in the inset where the $q^{(2)}_z(t)$ data at the two lowest temperatures 
are shown.
In contrast to the Heisenberg case, a comparison of the
$L=16$ data with the $L=12$ data indicates that some amount of finite-size
effect still remains in the second term of eq.(5), and hence, in $q^{(2)}_z(t)$
itself. Nevertheless, the almost linear behavior without any
discernible up-bending tendency is robustly observed in common for $L=12$
and $L=16$, suggesting that this feature is a bulk property.

In fact, a similar linear behavior
was also observed in the 3D Ising SG
in zero field\cite{Ogielsky},
where the existence of a finite-temperature SG transition is
established\cite{Review}. Unfortunately, 
we cannot tell from \figtag\ref{fig_Ct_3DISG} whether a
true phase transition occurs or not in the 3D Ising SG in a field, 
since an apparently linear behavior could also arise
when the correlation time simply
exceeds the time window of the simulation,   but stays finite.
Yet, we may safely conclude here that the ordering behavior of the Ising SG
differs significantly from that of the weakly anisotropic Heisenberg SG,
despite the similarity in their global symmetry properties. 

\begin{figure}[ht]
  \begin{center}
    \includegraphics[width=\linewidth]{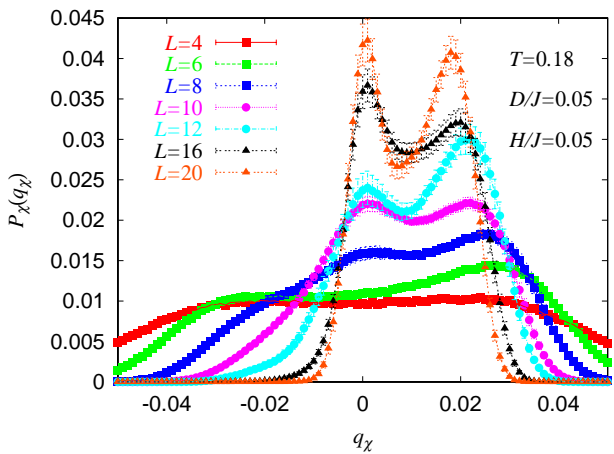}
\caption{The  chiral-overlap distribution function
at a temperature $T=0.18$ for $D/J=0.05$ and $H/J=0.05$.
The transition temperature is 
$T_g \simeq 0.21$.}
    \label{fig_Pqx}
  \end{center}
\end{figure}

In order to get further insights
into the nature of the RSB transition of the present model,
we show in \figtag\ref{fig_Pqx} the chiral-overlap distribution function $P(q'_\chi)\equiv
[\thermav{\delta(q_\chi-q'_\chi)}]$ at $T=0.18$,
somewhat below $T_g$. 
In addition to the primary peak
corresponding to $q_{\chi}=q_{\chi}^{\rm EA}>0$,
which grows and sharpens with increasing $L$,
there appears the second peak at $q_{\chi}\simeq 0$,
which also grows and sharpens with increasing $L$.
The existence of two distinct
peaks, both growing and sharpening with increasing $L$,
is a clear indication of the occurrence of RSB.
As observed in Ref.\cite{HukuKawa}, 
$P_{\chi}(q_{\chi})$  in zero field exhibits a feature of 
one-step-like RSB, 
a central peak at $q_\chi=0$  
coexisting with the 
self-overlap  peak at $q_\chi=q_\chi^{{\rm EA}}$.  
The $P_{\chi}(q_{\chi})$ observed  here may be regarded
as the in-field counterpart of the zero-field $P_{\chi}(q_{\chi})$,
with a feature of such one-step-like RSB.

%
%
%
%

%
%

In summary,  by performing a large-scale equilibrium
MC simulation, we have shown that 
the weakly anisotropic 3D Heisenberg SG in magnetic fields
exhibits a thermodynamic  RSB transition in the chiral sector,
which accompanies the 
simultaneous SG order.
The ordering behavior of the weakly anisotropic Heisenberg SG
differs significantly from that of the 
Ising SG despite the similarity in their global symmetry properties.
Our observation is fully
consistent with the spin-chirality decoupling-recoupling scenario, 
and might give a clue to resolve
the apparent experimental discrepancy between 
the strongly anisotropic Ising-like SG 
and the weakly anisotropic Heisenberg-like SG.

The numerical calculation was performed on the HITACHI
SR8000 at the supercomputer system, ISSP, University of Tokyo,
and Pentium IV clustering machines in our laboratory.
The authors are thankful to H. Yoshino and K. Hukushima for
useful discussions.

\end{document}